\documentclass[twocolumn,showpacs,prc]{revtex4}

\usepackage{graphicx}
\usepackage{dcolumn}
\usepackage{bm}

\newcommand {\nc} {\newcommand}
\nc {\beq} {\begin{eqnarray}}
\nc {\eeq} {\end{eqnarray}}
\nc {\eeqn} [1] {\label{#1} \end{eqnarray}}
\nc {\eol} {\nonumber \\}
\nc {\eoln} [1] {\label{#1} \\}

\begin{document}

\preprint{TRI-PP-03-25}

\title{On the (absence of a) relationship between bound and scattering states
in quantum mechanics. Application to $^{12}$C + $\alpha$}
\author{Jean-Marc Sparenberg}\email{jmspar@triumf.ca}
\affiliation{Theory Group, TRIUMF, 4004 Wesbrook Mall,
Vancouver, BC, Canada V6T 2A3}
\date{\today}

\begin{abstract}
Using phase-equivalent supersymmetric partner potentials,
a general result from the inverse problem in quantum scattering theory
is illustrated,
i.e., that bound-state properties cannot be extracted from the phase
shifts of a single partial wave, as a matter of principle.
In particular, recent R-matrix analyses of the $^{12}$C + $\alpha$ system,
extracting the asymptotic normalization constant
of the $2^+$ subthreshold state, $C_{12}$,
from the $\ell=2$ elastic-scattering phase shifts and bound-state energy,
are shown to be unreliable.
In contrast, this important constant in nuclear astrophysics
can be deduced from the simultaneous analysis
of the $\ell=0$, 2, 4, 6 partial waves in a simplified potential model.
A new supersymmetric inversion potential and existing models give
$C_{12}=144.5\pm8.5 \times 10^3$ fm$^{-1/2}$.
\end{abstract}

\pacs{03.65.Nk,25.55.Ci,26.20.+f}

\maketitle

\section{Introduction}

Can bound-state properties be deduced from scattering data?
The answer to this question showed interesting evolutions throughout
the history of quantum scattering theory
(see Foreword by R.~G.\ Newton in Ref.\ \cite{chadan:77}),
starting with Heisenberg's conjecture in the early forties 
that the scattering (S) matrix contains all physical information,
including that regarding bound states.
This conjecture was later seriously weakened in the framework of the
potential model:
in 1949, Bargmann \cite{bargmann:49a,bargmann:49b} constructed potentials
that share the same S matrix for partial wave $\ell=0$
(so-called {\em phase-equivalent} potentials because they share the same
scattering phase shifts) but have different bound-state properties
[different energies, or identical energies but different wave functions].
On the other hand, that same year, Levinson
(see for instance Ref.\ \cite{chadan:77}) proved that,
for a given partial wave $\ell$,
the number of bound states, $N_\ell$, can be deduced from the phase shifts.
These results were then explained by the inverse-problem theory
\cite{chadan:77}, which shows that for a given partial wave
the potential is uniquely determined by
(i) the scattering phase shifts at all energies,
(ii) the bound-state energies
and (iii) one additional real parameter for each bound-state wave function
[e.g., the asymptotic normalization constant (ANC),
see definition (\ref{eq:ANCdef})].
This only applies to local central potentials displaying no singularity.
Levinson's theorem, however, was generalized by Swan in 1963 \cite{swan:63},
who proved that $N_\ell$ cannot be deduced from the S matrix
for potentials displaying an $r^{-2}$ repulsive singularity at the origin.
Since then,
the construction of phase-equivalent potentials for a given
partial wave has known important progress
\cite{baye:87a,baye:87b,ancarani:92,baye:93a,baye:94}
thanks to the use
of the algebraic formalism of supersymmetric quantum mechanics
\cite{witten:81,sukumar:85b,sukumar:85c}
used below (see Fig.~\ref{fig:potANC}).
%The whole family of potentials phase-equivalent to a given one 
%but with an arbitrarily-different bound spectrum
%is now expressed in a closed algebraic form \cite{baye:94}.

As far as only one partial wave is concerned, the answer to the above question
is thus rather clear-cut:
neither the number of bound states, nor their energies nor their ANCs 
can be deduced from the S matrix alone.
Moreover, these bound-state properties are independent of one another,
which means that knowing, for instance,
both the S matrix and a bound-state energy
does not constrain the value of the ANC as long as no other physical
information is available on this particular bound state
(e.g., its radius or its lifetime) or on the considered system in general
(e.g., the range of the interaction).
This impossibility proof being established
in the framework of the potential model,
which is a particular case of more general models
(e.g., microscopic or phenomenological models),
it holds for these more general models too.
However, several works
\cite{iwinski:84,blokhintsev:84,blokhintsev:93,angulo:00,tischhauser:02} 
seem to ignore this argument and aim at extracting the ANC of a bound-state
with known energy from the corresponding partial-wave S matrix.
In Refs.\ \cite{angulo:00,tischhauser:02} for instance,
the reaction- (R-)matrix formalism \cite{wigner:47,lane:58}
is used to extract the ANC of an $^{16}$O bound state
from $^{12}$C + $\alpha$ elastic-scattering data.
In the R-matrix phenomenological model,
it is rather natural to assume that bound states have a measurable impact
on the partial-wave S matrix, like elastic-scattering resonances,
because both bound and resonant states are characterized by an energy 
and a reduced width.
However, in scattering theory,
bound and resonant states have very different natures,
and the above impossibility argument seriously questions
the ANC extraction of Refs.\ \cite{angulo:00,tischhauser:02}.
In Sec.\ \ref{sec:R-matrix}, I show how that misuse of the
R-matrix formalism actually explains the inconsistency between
the ANC values obtained in these references.
Then in Sec.\ \ref{sec:SUSY}, I explicitly construct simple potential models
which illustrate the independence of the bound-state ANC from the phase shifts
for a single partial wave.

Now, considering several partial waves simultaneously leads to a less clear
situation.
The inverse-problem theory implies that a regular interaction potential
is uniquely determined by phase shifts of a given partial wave with angular
momentum high enough so that the partial wave contains no bound state.
Using inversion techniques \cite{chadan:77},
a unique potential can thus in principle be deduced from such phase shifts;
this potential can in turn be used for lower partial waves, for which it may
sustain bound states.
In that case, bound state properties (of low partial waves) are totally
determined by scattering phase shifts (of a high partial wave).
However, this program presents three difficulties.
First, it assumes that the potential is local and energy independent,
which, in general,
is not a realistic approximation when the interacting particles
have an internal structure (see the discussion in Secs.\ \ref{sec:inversion}
and \ref{sec:conclusion} for the particular case of $^{12}$C + $\alpha$).
Second, it requires that the interaction be independent of the partial wave,
which is not always the case:
scattering phase shifts of several partial waves are not
necessarily compatible with one unique potential.
Third, it is based on the knowledge of the phase shifts of a
high-angular-momentum partial wave at {\em all energies},
which are generally not available:
at high energy, the complexity of multichannel effects increases,
which makes the data difficult to analyze;
and at low energy, the phase shifts of high-angular-momentum partial waves
are very small because of the centrifugal barrier.
Hence, considering several partial waves at the same time may help
deducing bound-state properties from scattering data,
but the practical applicability of the method could strongly depend on the
considered system.
In Sec.\ \ref{sec:inversion},
I show that for the particular case of $^{12}$C + $\alpha$
encouraging results are obtained
by considering several partial waves at the same time,
in the framework of a simplified model
in which $^{12}$C and $\alpha$ are considered as rigid
nuclei in their $0^+$ ground state, interacting through a local energy-
and angular-momentum-independent potential.
These results are then compared with more sophisticated models available
in the literature.
Conclusions and perspectives for the $^{12}$C + $\alpha$ system are given
in Sec.\ \ref{sec:conclusion}.

\section{\label{sec:R-matrix}
R-matrix analyses}

The $^{12}$C + $\alpha$ system is particularly important
in nuclear astrophysics since the
$^{12}$C ($\alpha$, $\gamma$) $^{16}$O capture is a key reaction
in the helium-burning phase of red giant stars:
by competing with the triple $\alpha$ reaction (leading to $^{12}$C),
it determines the ratio of carbon and oxygen which results from
stellar nucleosynthesis.
Experimentally, the Coulomb repulsion between $^{12}$C and
$\alpha$ makes the direct measurement of the capture reaction impossible
at the very low energies of astrophysical interest (around 300 keV
in the center-of-mass frame).
Hence, available experimental results at higher energies
($E_\mathrm{c.m.}>1$ MeV) have to be extrapolated to these low energies.
This extrapolation,
generally performed with the R-matrix model
(the only model to date able to precisely fit all the available data),
is itself problematic because of the presence of
two $^{16}$O bound states just below the $^{12}$C + $\alpha$ threshold.
Whereas the influence of the $E_\mathrm{c.m.}=-45.1\pm0.1$ keV state is
well understood \cite{azuma:94,brune:99},
the influence of the $E_\mathrm{c.m.}=-244.9\pm0.6$ keV state
is still the major source of uncertainty on the reaction rate
\cite{brune:99,angulo:00,tischhauser:02}.
The unknown key quantity is its ANC
(the energies of both states are well known),
which is related to its $\alpha$ reduced width in the R-matrix formalism
(I do not use reduced widths here since they depend on the R-matrix radius,
which makes comparison between different models difficult).

The $-245$ keV subthreshold state has a positive parity
and a $J=2$ angular momentum ($2^+$ state);
microscopic cluster models (see Ref.\ \cite{descouvemont:87} and references
therein) suggest that its structure is dominated by
a $^{12}$C cluster and an $\alpha$ cluster,
both in their fundamental $0^+$ state,
with a relative motion of angular momentum $\ell=2$.
It is thus on the phase shifts of this partial wave
that one could naively expect
to see the influence of the $2^+$ state,
and it is from these phase shifts that one could try to extract its ANC,
$C_{12}$ (where the second index refers to $\ell=2$ and the
first index means that this state is the lowest one for this partial wave).
This constant is defined by the asymptotic behavior of the
radial part, $u_{12}(r)$, of the relative wave function
between the $^{12}$C and $\alpha$ nuclei for this particular bound state:
\beq
u_{12}(r) \mathop{\sim}_{r \rightarrow \infty}
C_{12} W_{-6.615, 5/2} (0.3751 \; r),
\eeqn{eq:ANCdef}
where $r$ is the radial coordinate (in fm) and $W$ is the Whittaker function
\cite{abramowitz:65}
(a decreasing exponential function in the absence of Coulomb interaction).

In Ref.\ \cite{angulo:00}, the $\ell=2$ phase shifts from Ref.\ \cite{plaga:87}
are analyzed by the R-matrix model,
leading to a minimum in the $\chi^2$ (see Fig.\ 1 of Ref.\ \cite{angulo:00})
for $C_{12}=402 \times 10^3$ fm$^{-1/2}$.
However, this extraction is very delicate since the influence of the
subthreshold state has to be disentangled from the background phase shifts
in the energy region where experimental data are available;
this influence is smaller (see Fig.\ 2 of Ref.\ \cite{angulo:00}) than
the main components of this background,
i.e., the hard-sphere and high-energy-pole phase shifts.
As a consequence, this minimum strongly depends on the range of data fitted
(see Fig.\ 1 of Ref.\ \cite{angulo:00}).
Moreover, I have checked that the minimum is highly sensitive to 
the high-energy background-pole energy,
which is fixed at 10 MeV in Ref.\ \cite{angulo:00}.
Without this restriction, very different values of $C_{12}$ are obtained;
hence, the error on $C_{12}$, which is not estimated in Ref.\ \cite{angulo:00},
is actually very large.

In Ref.\ \cite{tischhauser:02},
the $^{12}$C + $\alpha$ elastic scattering has been remeasured with very high
precision.
An R-matrix analysis with free high-energy-pole energies
leads to a $\chi^2$ minimum (see Fig.\ 2 (a) of Ref.\ \cite{tischhauser:02})
for $C_{12}=154 \pm 18 \times 10^3$ fm$^{-1/2}$.
The error bar on $C_{12}$ is estimated by following the guideline
\cite{press:89,azuma:94}
\beq
\chi^2 < \chi^2_\mathrm{min} + 9 \chi^2_\mathrm{min}/\nu,
\label{eq:chi2}
\eeq
where $\chi^2_\mathrm{min}$ is the minimal $\chi^2$ and $\nu$
is the number of degrees of freedom.
In the present case, one has $\chi^2_\mathrm{min}=18941$
(the $\chi^2$ of Fig.\ 2 in Ref.\ \cite{tischhauser:02} have to be divided
by the 32 reference angles),
for 11392 data points and 65 parameters
(32 R-matrix parameters and 33 experimental parameters),
which leads to $\nu=11327$.
Following the theory of $\chi^2$ minimization \cite{press:89},
one finds that the probability of having $\chi^2\ge 18941$ for
$\nu=11327$ is of the order of $10^{-391}$,
which probably indicates that some systematic errors have been underestimated
in Ref.\ \cite{tischhauser:02}.
Consequently, Eq.\ (\ref{eq:chi2}) cannot be used
to define error bars on parameters since this equation only takes into account
statistical errors and requires that the fit be good
in the sense of $\chi^2$ theory.
A less rigorous way of checking the validity of a fit is to use the
``chi-by-eye'' approach \cite{press:89}.
In the case of Ref.\ \cite{tischhauser:02},
the fit looks good (see Fig.\ 1 of this reference),
which is related to the fact that $\chi^2_\mathrm{min}/\nu=1.672$
is not that far from 1.
However, the fits obtained with $C_{12}=136$ or $172 \times 10^3$ fm$^{-1/2}$,
which correspond to $\chi^2/\nu=1.674$, look probably equally good.
I thus consider that the error bar on $C_{12}$ obtained in Ref.\
\cite{tischhauser:02} is underestimated.

In summary, the values of $C_{12}$ obtained in both Refs.\
\cite{angulo:00} and \cite{tischhauser:02} are most probably  unreliable,
for different reasons.
This explains why,
despite the fact that the $\ell=2$ phase shifts of Refs.\ \cite{plaga:87}
and \cite{tischhauser:02,buchmann:03} are compatible with one another,
their R-matrix analyses give totally incompatible results for $C_{12}$.
One might think that this inconsistency is due to the fact that
in Ref.\ \cite{tischhauser:02} all relevant partial waves are taken into
account whereas in Ref.\ \cite{angulo:00} only $\ell=2$ is considered.
However, since the R-matrix formalism is based on a
partial-wave decomposition of the cross section,
$C_{12}$ is actually independent of the $\ell\ne2$ parameters
(in Ref.\ \cite{tischhauser:02},
the only influence of the $\ell\ne2$ phase shifts on $\ell=2$
comes through the choice of a common R-matrix radius for all partial waves;
however, this does not seem to strongly constrain the fit of the $\ell=2$
phase shifts, as discussed in the next paragraph).
%In both analyses, the value of $C_{12}$ is thus deduced from the $\ell=2$
%phase shifts only.
I argue that this incompatibility rather illustrates the impossibility,
well-known in quantum scattering theory,
of deducing a bound-state ANC when only the bound-state energy and
the elastic-scattering phase shifts of the corresponding partial wave
are known.
% as discussed in the above paragraph:
%In principle, a subthreshold state should not improve the quality
%of an R-matrix fit of elastic scattering data
%and an R-matrix fit of the $\ell=2$ phase shifts
%with fixed bound-state energy should not give any constraint on $C_{12}$.

Let us however notice that, in the case of the R-matrix phenomenological model,
this impossibility is probably not as strong as in general scattering theory.
As shown in the next Section (see Fig.\ \ref{fig:potANC}),
the range of a $^{12}$C + $\alpha$ interaction that reproduces
both the $\ell=2$ phase shifts and the bound-state energy increases with the
value of $C_{12}$.
But the R-matrix formalism is only valid when the nuclear interaction is
negligible above the R-matrix radius.
In the R-matrix {\em theory},
where the number of R-matrix poles is infinite \cite{lane:58},
this does not cause any problem:
the results are independent of the R-matrix radius,
which can thus be chosen as large as necessary
(in other words, it would be possible to calculate an R-matrix for any of the
potentials plotted in Fig.\ \ref{fig:potANC} but the R-matrix radius should be
larger than the potential range). 
In the R-matrix {\em phenomenological model},
where the number of poles is limited to the number of bound states and
resonances in the energy interval considered,
plus one possible high-energy pole to help simulating the background,
a constraint appears on the R-matrix radius:
in Ref.\ \cite{vogt:96}, a value of 5.5-6 fm is found,
based on the $\ell=1$ partial wave;
in Ref.\ \cite{tischhauser:02} [see Fig.\ 2 (b)] 5.5 fm leads to the best fit
(I have checked that this constraint is mainly due
to the $\ell=0$ partial wave, for which the background is maximal);
in Ref.\ \cite{angulo:00}, the radius is fixed at 6.5 fm.
Hence, very large values of $C_{12}$ can probably be excluded by an
R-matrix fit; very small values, on the other hand, cannot be excluded.
I could actually find very good fits of the $\ell=2$ phase shifts
with $C_{12}=0$ for R-matrix radii in the range 4.5-6.5 fm.
This confirms the doubts expressed above on the analyses of Refs.\
\cite{angulo:00,tischhauser:02} and it shows that, in this particular case,
the quality of the fit does not strongly depend
on the precise value of the R-matrix radius.
\begin{figure}
\scalebox{0.45}{\includegraphics{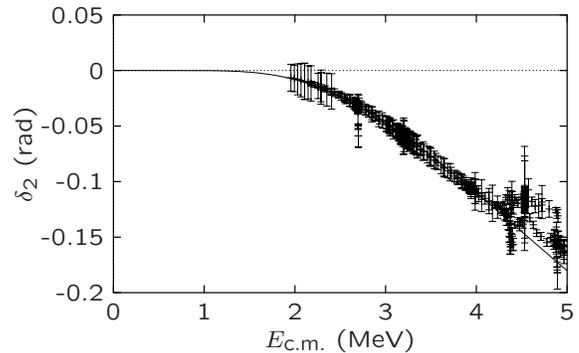}}
\caption{\label{fig:d2}
Experimental $\ell=2$ background phase shifts deduced
from the $^{12}$C + $\alpha$ elastic-scattering data of
Ref.\ \cite{tischhauser:02,buchmann:03} (points),
compared with the theoretical phase shifts of the phase-equivalent potentials
of Fig.\ \ref{fig:potANC} (solid line, identical for all potentials).}
\end{figure}

\section{\label{sec:SUSY}
Inversion of one partial wave}

\begin{figure}
\scalebox{0.45}{\includegraphics{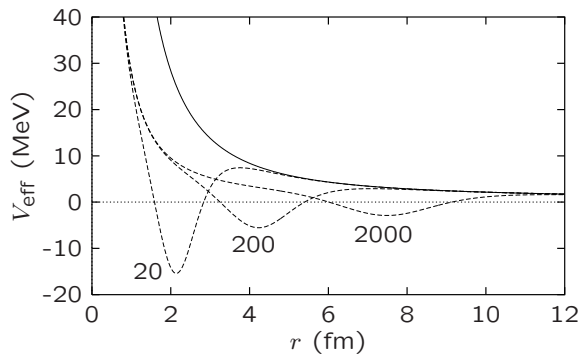}}
\caption{\label{fig:potANC}
Effective $^{12}$C + $\alpha$ interaction potentials for $\ell=2$.
The solid-line potential corresponds to Eq.\ (\ref{eq:pot})
while the dashed-line potentials are its phase-equivalent supersymmetric
partners with one bound state at $-245$ keV,
for $C_{12}=20$, 200 and $2\,000 \times 10^3$ fm$^{-1/2}$.}
\end{figure}

Up to now, the discussion is based on general arguments from scattering theory.
Let me now construct simplified two-body potential models
that illustrate these arguments in the particular case of $^{12}$C + $\alpha$,
using the supersymmetric-quantum-mechanics
\cite{witten:81,sukumar:85b,sukumar:85c} inversion technique of 
Refs.\ \cite{sparenberg:97a,sparenberg:00c}.
Starting from a phase-shift analysis \cite{buchmann:03} 
of the purely elastic data
($E_\mathrm{c.m.} < 4.966$ MeV) of Ref.\ \cite{tischhauser:02},
I have removed, using the R-matrix formalism,
resonances that have a more complicated structure
than $^{12}$C + $\alpha$ in their $0^+$ ground states
(such resonances cannot be explained by the present simplified model).
For $\ell=2$, this means removing the narrow resonance at 2.683 MeV and
the wide resonance at 4.358 MeV \cite{tilley:93},
which leads to the phase shifts plotted in Fig.\ \ref{fig:d2}
(the spread of the data above 4.2 MeV is due to the difficulty of performing a
reliable phase-shift analysis in the vicinity of a resonance).
By solving the so-called ``singular inverse problem'' defined in Ref.\
\cite{sparenberg:97a},
I have constructed a potential that reproduces these phase shifts with
very high precision.
This potential has an $r^{-2}$ singularity at the origin and satisfies
Swan's theorem \cite{swan:63};
to simplify further developments,
its nuclear part has been approximated by the analytical expression
\beq
V(r)=V \exp(-r^2/R^2)/r^2,
\eeqn{eq:pot}
where $V=43.40$ MeV fm$^2$ and $R=5.091$ fm.
The corresponding $\ell=2$ effective potential (centrifugal barrier plus
Coulomb and nuclear interactions) is represented in Fig.\ \ref{fig:potANC}
(solid line) and its phase shifts are shown in Fig.\ \ref{fig:d2}.
Surprisingly, a potential of the form (\ref{eq:pot}) is also able 
to reproduce the $\ell=0$ background phase shifts with satisfactory precision,
which suggests an alternative way of parameterizing
the $^{12}$C + $\alpha$ background in future R-matrix analyses
(in the spirit of the ``hybrid'' R-matrix model of Refs.\
\cite{johnson:73,koonin:74,langanke:85}).
The fit of the $\ell=2$ scattering phase shifts is very good,
despite the fact that potential (\ref{eq:pot}) has, by construction,
{\em no bound state} (it is purely repulsive);
this is a first proof that, from the point of view of scattering theory,
the $\ell=2$ background phase shifts are totally disconnected
from the $2^+$ subthreshold bound state.

\begin{figure}
\scalebox{0.45}{\includegraphics{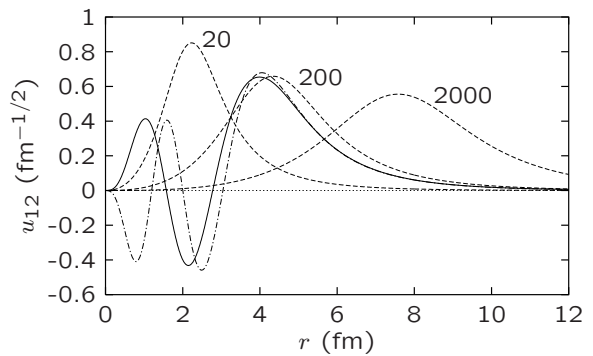}}
\caption{\label{fig:bs}
Radial part of the $2^+$ subthreshold state
$^{12}$C + $\alpha$ relative wave functions.
The dashed lines correspond to the potentials of Fig.\ \ref{fig:potANC},
the solid line to the deep inversion potential and the dash-dotted line to the
potential of Ref.\ \cite{buck:84}.}
\end{figure}

Now, since this state is known to exist and to have a structure such that
it could be well simulated by the present simplified model,
it is physically meaningful to perform a further step in the inversion of the
$\ell=2$ phase shifts,
i.e., to add a bound state to the above potential without modifying its
phase shifts \cite{sparenberg:00c}.
This phase-equivalent bound-state addition may be carried out
through a further use of supersymmetric quantum mechanics \cite{baye:87b};
it introduces two arbitrary parameters: the bound-state energy and its ANC.
In the present case, the energy is known experimentally but not the ANC;
hence, there is a one-parameter family of $\ell=2$ effective potentials
that share identical phase shifts (Fig.\ \ref{fig:d2})
and have a bound-state at $-245$ keV.
Such potentials are represented by dashed lines for three arbitrary values of
$C_{12}$ in Fig.\ \ref{fig:potANC};
they constitute generalizations (to $\ell \ne 0$ and in the presence of a 
Coulomb interaction) of Bargmann's potentials
(see Fig.\ 1 in Ref.\ \cite{bargmann:49b}).
The corresponding bound-state wave functions are plotted in Fig.\ \ref{fig:bs}
(dashed lines),
where their very different asymptotic behaviors are clearly seen.
This proves that,
from the $\ell=2$ phase shifts and bound-state energy only,
there is no way of telling what is the best value for $C_{12}$,
in contrast with R-matrix attempts described above
(see however the discussion of the R-matrix phenomenological model in the
previous Section).

\section{\label{sec:inversion}
Inversion of several partial waves}

The general discussion from the introduction suggests to look at other partial
waves simultaneously.
The $2^+$ subthreshold state is actually a member of a rotational band,
together with the $0^+$ bound state at $-1\,113\pm1$ keV
and the $4^+$ and $6^+$ resonances at $3\,194\pm3$ keV (width: $26 \pm 3$ keV)
and $9\,113\pm7$ keV (width: $420\pm20$ keV) respectively \cite{tilley:93}.
This is indicated by cluster-model calculations
(see Ref.\ \cite{descouvemont:87} and references therein),
which give a fair description of these four states in a single-channel
calculation where both the $^{12}$C and $\alpha$ nuclei are
in their $0^+$ ground state.
This suggests to ask the following question:
could all these states be reproduced by a unique $\ell$-independent potential?
Of course, in principle,
the internal structure of the nuclei could play a role and a satisfactory
potential could be non local and energy dependent.
It is however worth testing the simplest hypothesis, i.e.,
that of an energy- and $\ell$-independent local potential,
because if such a potential were sufficient,
it would be strongly constrained by the data of the four partial waves
considered simultaneously,
which could help solving the $C_{12}$ indefiniteness.

I could not find such a potential
within the above one-parameter potential family.
This is not very surprising since general arguments
from the microscopic cluster model show \cite{friedrich:81}
that $\ell$-independent interaction potentials between nuclei should have
low-energy non-physical bound states
that simulate the Pauli principle between the constituting nucleons
[because of these so-called {\em Pauli forbidden states} (PFSs),
such potentials are {\em deep},
in contrast with {\em shallow} potentials
that only sustain physical bound states].
For the $\ell=2$ partial wave of $^{12}$C + $\alpha$,
the expected number of PFSs is two \cite{langanke:85}.

In the context of inversion,
such bound states complicate things a lot since each of them introduces two
additional arbitrary parameters (neither their energy nor their ANC
is known experimentally).
Following Ref.\ \cite{sparenberg:00c}, 
I introduce these states phenomenologically by regularizing the above
inversion potential with the expression $V_\mathrm{reg} + a r^2 + b r^3$
for $r\le r_\mathrm{reg}$,
where $V_\mathrm{reg}$ and $r_\mathrm{reg}$ are fitting parameters 
($a$ and $b$ are calculated to make the potential and its first derivative
continuous in $r_\mathrm{reg}$).
Since this regularization slightly alters the phase shifts,
parameters $V$ and $R$ in Eq.\ (\ref{eq:pot}) have to be adapted
to maintain the fit of Fig.\ \ref{fig:d2}.
Altogether, there are thus five parameters ($V$, $R$, $C_{12}$,
$V_\mathrm{reg}$ and $r_\mathrm{reg}$) which can be varied to fit all the
$\ell=0,2,4,6$ data at the same time.
The best fit I have obtained is for $V=63.81$ MeV fm$^2$,
$R=4.925$ fm, $V_\mathrm{reg}=-107.9241$ MeV, $r_\mathrm{reg}=4.1$ fm.
By construction, this inversion potential perfectly reproduces
the $2^+$ state energy.
The wave function is represented in Fig.\ \ref{fig:bs} (solid line);
its two nodes are due to its orthogonality to the PFSs at $-46$ and $-18$ MeV,
which confirms that the potential is deep.
The corresponding value for $C_{12}$ is $137 \times 10^3$ fm$^{-1/2}$.

I consider this prediction as reliable because
all the even phase shifts in the elastic region are also rather well fitted
(for example, the $4^+$ resonance is at $3\,196$ keV with width 22 keV).
However, this potential underestimates the $0^+$ bound-state energy
($-1\,696$ keV), as well as the $6^+$ resonance energy ($8\,985$ keV)
and width (259 keV).
These inaccuracies are probably due to the extreme simplicity of the
present model (energy- and $\ell$-independent local potential) and it is
worth comparing it with more sophisticated models of the literature,
which better take into account many-body effects.
The two non-local $\ell$-dependent potentials of Ref.\ \cite{langanke:85},
which also fit the physical properties of the members of the
rotational band,
lead to similar wave functions for the 2$^+$ subthreshold state (2 nodes).
The Gaussian potential provides $C_{12}=144 \times 10^3$ fm$^{-1/2}$
while the Woods-Saxon potential provides $153 \times 10^3$ fm$^{-1/2}$.
In Ref.\ \cite{buck:84}, a simple local potential is given,
which precisely reproduces the energies of the members of the rotational band,
as well as the resonance widths,
provided a slight $\ell$ dependence is allowed.
Although the number of PFSs is three for this potential,
which adds a node in the subthreshold-bound-state wave function 
(see dash-dotted line in Fig.\ \ref{fig:bs}),
it provides a value of $C_{12}$ very similar to the inversion potential:
$136 \times 10^3$ fm$^{-1/2}$.

\section{\label{sec:conclusion}
Conclusion}

I have illustrated on $^{12}$C + $\alpha$ two general results from
the inverse problem in quantum scattering theory:
(i) for a given partial wave, the scattering phase shifts,
the number of bound states, their energies and their ANCs
are independent of one another;
(ii) when several partial waves can be described by the same theoretical model,
these quantities may be related to one another.
This is reasonably the case for $^{12}$C + $\alpha$,
for which I have deduced the ANC value of the $2^+$ subthreshold state from the
simultaneous analysis of the $\ell=0,2,4,6$ partial waves,
with the help of a local energy- and angular-momentum-independent potential
constructed by supersymmetric inversion.

Since this potential lies on rather restrictive hypotheses and is not able
to exactly fit all the data,
I have compared it with results from more sophisticated potential models
available in the literature.
All theoretical predictions lie in a rather limited range,
$C_{12}=144.5 \pm 8.5 \times 10^3$ fm$^{-1/2}$,
which confirms somehow their overall validity.
The error bar can thus be considered as a sensible estimate of the theoretical
uncertainty on $C_{12}$ in a $^{12}$C + $\alpha$ potential model.
The fact that both local (present work and Ref.\ \cite{buck:84})
and non-local \cite{langanke:85} potentials give consistent results is an
indication that non-local effects are not very important for this particular
state.
This confirms that its structure is dominated by rigid $^{12}$C and $\alpha$
clusters with $\ell=2$ relative angular momentum.
A similar smallness of non-local effects has been found in Ref.\
\cite{baye:85} for the $^{16}$O + $\alpha$ system.

All the potential models discussed above neglect multichannel effects,
which means that the predicted value of $C_{12}$ might be overestimated.
This would be consistent with the experimental value obtained
from transfer reactions \cite{brune:99},
$C_{12}=114 \pm 10 \times 10^3$ fm$^{-1/2}$,
and with the theoretical estimate from the microscopic multiconfiguration
cluster model of Ref.\ \cite{descouvemont:87},
$C_{12}=134 \times 10^3$ fm$^{-1/2}$.
Whereas compatible with the unreliable value from Ref.\ \cite{tischhauser:02},
the above theoretical estimate excludes that from Ref.\ \cite{angulo:00},
as well as the cascade-transition value
$C_{12}=228^{+33}_{-37} \times 10^3$ fm$^{-1/2}$ from Ref.\ \cite{buchmann:01}.
A new measurement of the cascade transitions is planned at TRIUMF,
which hopefully should clarify the situation.
In the meanwhile, R-matrix fits should reflect the uncertainty on $C_{12}$.

Let me end with a general discussion of possible theoretical models for
$^{12}$C + $\alpha$ and similar systems relevant to nuclear astrophysics.
Ideally, {\em ab initio} calculations should be performed;
however, despite the success of the microscopic cluster model,
such calculations are still too primitive to provide precise predictions of
all available data.
Hence, phenomenological models still have an important role to play.
Among them, the potential model is probably the most physical and intuitive
one.
However, simulating the full complexity of the many-body nuclear system
requires a non-local \cite{langanke:85} multichannel \cite{baldock:84}
treatment.
This considerably complicates the model and a good quality fit
of all available data with the potential model is not available to date.
In contrast, the R-matrix model allows such a fit because it remains rather
simple even in the presence of many channels.
The price to pay for that is a large number of parameters and a very
phenomenological description, in particular of background terms.
An interesting approach which combines the advantages of both the potential
and R-matrix models is that of ``hybrid'' models
\cite{johnson:73,koonin:74,langanke:85}.
The potentials constructed in the present work renew the interest for these
models.
In particular, purely repulsive potentials of the form (\ref{eq:pot}) could be
used as a new description of the background,
a possibility which will be explored in a future work.

\begin{acknowledgments}
I acknowledge very useful discussions with C.\ Barbieri, F.\ Barker, D.\ Baye,
L.\ Buchmann, P.\ Descouvemont, A.\ Mukhamedzhanov and B.\ Jennings.
I also thank L.\ Trache and C.\ Brune for drawing my attention to Refs.\
\cite{brune:99} and \cite{iwinski:84} respectively.
\end{acknowledgments}

%\bibliographystyle{apsrev}
%\bibliography{$HOME/Biblio/own,$HOME/Biblio/others}
%\end{document}

\end{document}